%% file: main.tex
\documentclass[]{meta}

\input{shared-preamble}

\newcommand{\isarxiv}{}

\begin{document}

\title{REAP: Automatic Curation of Coding Agent Benchmarks from Interactive Production Usage}
\author[*]{Smriti Jha}
\author[*]{Matteo Paltenghi}
\author{Chandra Maddila}
\author{Vijayaraghavan Murali}
\author{Shubham Ugare}
\author{Satish Chandra}
\affiliation{Meta, USA}
\contribution[*]{Equal contribution}
\correspondence{\texttt{\{smrj,mattepalte\}@meta.com}}
\metadata[Code]{\url{https://anonymous.4open.science/r/REAP_artifact}}
\abstract{\input{abstract}}
{%
  \setlength{\parskip}{0pt}%
  \noindent\small Accepted at ASE 2026 Industry Showcase\par
  \vspace{-6pt}\noindent\rule{\textwidth}{0.4pt}\par
}
\vspace{0.9cm}
\maketitle
\input{content}

\bibliographystyle{ACM-Reference-Format}
\bibliography{references,MatteoAtMeta}

\end{document}

%% file: shared-preamble.tex
\usepackage{multirow}
\usepackage{tikz}
\usepackage{enumitem}
\setlist[itemize]{leftmargin=*, nosep}
\setlist[enumerate]{leftmargin=*, nosep}
\usepackage{pifont}

\usetikzlibrary{shapes.geometric, arrows.meta, positioning, calc, shadows.blur, fit, backgrounds}
\usepackage{fontawesome5}

\input{macros}

\definecolor{insightbg}{HTML}{F1F4F7}
\newtcolorbox{insight}{
  colback=insightbg,
  colframe=insightbg,
  boxrule=0pt,
  arc=2mm,
  left=6pt,
  right=6pt,
  top=6pt,
  bottom=6pt,
  before upper={\textbf{Key Insight.}\enspace},
}

%% file: macros.tex
\newcommand{\benchmark}{Harvest}
\newcommand{\pipeline}{REAP}

\newcommand{\SolveRateHighest}{58.2}
\newcommand{\SolveRateLowest}{42.9}
\newcommand{\SolveRateBestModel}{Claude Opus 4.6}

\newcommand{\AuditAN}{108}
\newcommand{\AuditBN}{132}
\newcommand{\AuditOverlapN}{95}
\newcommand{\AuditRetentionPct}{88}
\newcommand{\AuditNetGainPct}{22}

\newcommand{\taskClassifTotal}{100}
\newcommand{\taskClassifConsensusTestable}{57}
\newcommand{\taskClassifConsensusNontestable}{43}

\newcommand{\taskClassifFP}{3}
\newcommand{\taskClassifFN}{10}
\newcommand{\taskClassifAgree}{87}

\newcommand{\taskClassifAcc}{87.00\%}
\newcommand{\taskClassifPrec}{94.00\%}
\newcommand{\taskClassifRec}{82.46\%}
\newcommand{\taskClassifFone}{87.85\%}

\newcommand{\taskClassifKappa}{0.74}
\newcommand{\taskClassifIAA}{85\%}
\newcommand{\taskClassifIAAKappa}{0.70}

\newcommand{\testRelevanceTotal}{50}
\newcommand{\testRelevanceDiffs}{37}

\newcommand{\testRelevanceFP}{3}
\newcommand{\testRelevanceFN}{4}
\newcommand{\testRelevanceAgree}{43}

\newcommand{\testRelevanceAcc}{86.00\%}
\newcommand{\testRelevancePrec}{88.00\%}
\newcommand{\testRelevanceRec}{84.62\%}
\newcommand{\testRelevanceFone}{86.27\%}

\newcommand{\testRelevanceKappa}{0.72}
\newcommand{\testRelevanceIAA}{70\%}
\newcommand{\testRelevanceIAAKappa}{0.41}

%% file: abstract.tex
Production deployment of AI coding agents requires fast, reproducible evaluation signals. Existing industrial practices trade off speed and fidelity: online A/B testing takes weeks and risks user experience, shadow deployment yields signals that are not reproducible across runs, and public benchmarks diverge from production workloads in language distribution, prompt style, and codebase structure. This paper presents \pipeline{} (Relevance and Execution-Audited Pipeline), an automated curation pipeline that constructs production-derived benchmarks from real developer-agent sessions without manual labeling.

Such curation, while in-distribution to production usage, runs into several challenges. Untestable prompts, misaligned tests, and test flakiness all compromise evaluation reliability. While tasks can be manually audited to ensure only high-quality tasks remain in the benchmark, this approach is infeasible in the monorepo setting: the build infrastructure state is often ephemeral in large monorepos and requires the benchmark to be continuously re-curated against the current codebase. As manual verification cannot be sustained at this cadence, \pipeline{} adds an automated verification layer using LLM-based task classification, agentic test-relevance validation, and multi-run stability checks to ensure the executable benchmark yields trustworthy signals.

We use \pipeline{} to curate \benchmark{}, a benchmark where each task feeds the coding agent a real developer prompt and verifies the resulting code change against fail-to-pass tests retrieved from production. \benchmark{}'s distribution spans more than four programming languages with a majority of tasks drawn from Hack, a meaningful divergence from public benchmarks which are skewed towards Python. Model and harness evaluations reveal that solve rates range from \SolveRateLowest{}\% to \SolveRateHighest{}\% across five frontier models, surfacing capability differences that inform concrete deployment decisions. The underlying corpus contains proprietary source code and cannot be released publicly; we share our automated curation methodology and case studies to enable other organizations to construct similar production-derived benchmarks at scale.

%% file: content.tex
\section{Introduction}
\label{sec:intro}

AI coding agents are increasingly deployed in production software development environments, yet evaluating these agents reliably remains challenging.
Organizations must frequently compare foundation models, validate changes to agent configurations, and assess infrastructure updates. These decisions require fast, reproducible evaluation signals, but the available evaluation methods involve trade-offs.

Online evaluation through A/B testing yields high-fidelity signals grounded in real interactions, but requires weeks to achieve statistical significance, consumes substantial engineering resources, and risks degrading user experience during experimentation.
Shadow deployment runs candidate agents in parallel with production systems without serving their outputs to users, avoiding user disruption. However, it introduces non-determinism, as model outputs and environment state vary across runs, undermining reproducibility.
Offline evaluation using existing benchmarks offers speed and reproducibility, but public benchmarks differ from production workloads in multiple ways.
SWE-Bench~\cite{jimenezSWEbenchCanLanguage2024}, for example, sources tasks from Python GitHub issues, which differ from industrial workloads along several axes:
\begin{itemize}
    \item \textbf{Language distribution}: predominantly Python versus polyglot codebases spanning Hack, C++, Kotlin, and more.
    \item \textbf{Prompt style}: structured issue descriptions written for human developers versus informal, underspecified requests typed directly into an AI coding assistant.
    \item \textbf{Codebase structure}: standalone open-source repositories versus large-scale industrial monorepos with ephemeral tooling and distributed build infrastructure.
    \item \textbf{Task-type mix}: issue-derived benchmarks skew toward bug fixes, whereas production traffic includes refactoring, migrations, and feature work in comparable proportions.
    \item \textbf{Data contamination}: frozen benchmarks face growing contamination risk as their tasks appear in model training data.
\end{itemize}

These observations motivate a benchmark derived directly from production usage that:
(1) preserves verbatim prompts rather than reconstructing them synthetically,
(2) reflects the actual distribution of programming languages and task types in the target deployment,
(3) provides stable, execution-based evaluation signals suitable for both rapid experimentation and reinforcement learning, and
(4) supports rolling re-curation so that tasks remain fresh, executable, and post-date model training cutoffs.

To meet these goals, this paper presents \pipeline{} (Relevance and Execution-Audited Pipeline), an automated curation pipeline that constructs executable benchmarks from real developer-agent sessions in production. To help ensure reliability, \pipeline{} applies multiple filtering stages: LLM-based task classification to identify testable coding tasks, test relevance validation to confirm tests exercise the changed code, and multi-run stability checks to exclude flaky tests. Because monorepo infrastructure is ephemeral and historical snapshots degrade over time, the benchmark must be periodically re-curated rather than frozen; automation is therefore not a convenience but a prerequisite.

Using \pipeline{}, we curate a benchmark called \benchmark{}, where each task pairs a verbatim developer prompt with fail-to-pass (F2P) tests that provide an automated correctness signal without requiring LLM-based judges. We run \pipeline{} against a single production AI coding assistant, restricting the source pool to single-turn conversations; the resulting snapshot spans more than four programming languages with a majority of tasks drawn from Hack. Both the single-agent and single-turn restrictions are deliberate case-study scope choices rather than pipeline-level constraints. Evaluation of five frontier models on \benchmark{} yields solve rates ranging from \SolveRateLowest{}\% to \SolveRateHighest{}\%, with \SolveRateBestModel{} achieving the highest performance.

This paper makes the following contributions:
\begin{itemize}
\item \textbf{\pipeline{},} an automated curation pipeline that constructs execution-based benchmarks from production developer-agent sessions using relevance filtering and execution-based verification.
\item \textbf{Empirical validation of \pipeline{}'s} automated classifiers against human judgments, demonstrating near-human accuracy on task classification and test relevance.
\item \textbf{Case studies on \benchmark{},} where model and harness evaluations surface actionable capability differences that inform deployment decisions.
\item \textbf{\pipeline{} artifacts:} classifier prompts and annotation protocols, enabling other organizations to apply \pipeline{} to their own production data.
\end{itemize}

\ifdefined\isarxiv\else
The benchmark dataset itself cannot be released publicly because it contains internal developer prompts and proprietary source code. The released artifacts listed above are described in full in the Data Availability Statement.
\fi

\section{Benchmark Goals and Challenges}
\label{sec:challenges}

\pipeline{} is designed around three goals: (1) reflect realistic developer prompts, (2) operate in a large industrial monorepo with rolling re-curation, and (3) provide a reliable execution-based evaluation signal. Each goal introduces specific challenges.

\paragraph{Realistic prompts.}
Evaluating agents on the kinds of prompts developers actually issue requires more than just sourcing prompts from production; it requires preserving the conversational register and implicit context of real requests, and binding each prompt to the specific code change it produced. Reconstructing prompts from diff metadata, issue templates, or commit messages loses the authentic register of a developer-agent interaction and breaks the one-to-one link between prompt and code change, allowing multiple plausible interpretations rather than evaluating against the developer's actual intent. \pipeline{} therefore demands a verified one-to-one mapping between an authentic developer-agent conversation and the diff that landed in response, so each prompt--diff pair faithfully represents what the developer asked for and what was accepted.

\paragraph{Should work in a monorepo (the ``time travel'' problem).}
Building a benchmark from a large-scale monorepo introduces infrastructure constraints that do not appear in small open-source repositories.
Unlike settings where one can check out an old commit and deterministically re-run historical tooling and tests, industrial monorepos depend on remote services (code search, semantic indexing, file navigation) that continuously index the latest codebase state. These services provide only limited historical lookback and cannot be rolled back to match an older revision, causing drift when evaluating past commits.
Similarly, command-line tools and compiler binaries are deployed through package managers with automatic expiration (e.g., 30--365 days), preventing long-term use of archived toolchains.
Build systems and type checkers frequently use precomputed repository-wide indices that expire within days to weeks, and regenerating them on older commits may fail due to incompatibilities with current tooling.
Finally, contextual signals such as CI logs and failure traces are typically retained only briefly, and strict deployment policies may prohibit executing outdated code against production-adjacent services.
Together, these constraints motivate a rolling benchmark design where samples are periodically refreshed rather than frozen indefinitely.

\paragraph{Reliable evaluation signal.}
The value of an execution-based benchmark depends on the trustworthiness of its evaluation signal.
First, many real requests are not directly testable via execution (e.g., documenta\-tion-only changes, code explanation, UI-only tweaks), so the curation pipeline must filter to tasks that can be meaningfully evaluated with tests.
Second, even for testable tasks, naively running the entire continuous integration (CI) suite per candidate is often computationally expensive in a monorepo.
One practical solution is test discovery and relevance filtering: selecting a small set of tests that are actually affected by the change and that produce stable fail-to-pass behavior, rather than relying on coincidental failures, flakiness, or unrelated regressions.

\section{\pipeline{}}
\label{sec:curation}

\pipeline{} is an automated curation pipeline that produces execution-based benchmarks for evaluating AI coding agents. The tasks are curated from real sessions with a production coding assistant.

Figure~\ref{fig:pipeline-overview} gives a high-level overview of the pipeline. \pipeline{} operates in two phases. \emph{Dataset construction} (\S\ref{subsec:construction}) sources authentic developer conversations from a production coding assistant, creates an evaluation environment, and retrieves candidate tests. \emph{Verification} (\S\ref{subsec:verification}) then applies three filters, ordered so that semantic checks narrow the candidate set before execution: prompt quality filters remove leakage, non-testable, and template prompts; a test relevance filter drops tests with no causal connection to the diff; and multi-run test validation confirms the fail-to-pass signal while discarding flaky results. The surviving tasks form the final benchmark.

\definecolor{pipeblu}{HTML}{1565C0}
\definecolor{pipeblubg}{HTML}{DCEEFB}
\definecolor{pipeorg}{HTML}{E65100}
\definecolor{pipeorgbg}{HTML}{FFF3E0}
\definecolor{pipegrn}{HTML}{2E7D32}
\definecolor{pipegrnbg}{HTML}{E8F5E9}
\definecolor{pipearr}{HTML}{9E9E9E}

\begin{figure*}[t]
\centering
\resizebox{\textwidth}{!}{%
\begin{tikzpicture}[
  >={Stealth[length=6pt, width=4pt]},
  every node/.style={font=\small},
  sarrow/.style={->, pipearr, line width=1.6pt},
  stagebox/.style={
    rounded corners=8pt, line width=1.2pt,
    inner xsep=10pt, inner ysep=8pt, align=left,
    minimum height=2.2cm,
  },
  stagenum/.style={
    circle, minimum size=12pt, font=\tiny\bfseries,
    inner sep=0pt, text=white,
  },
]
\def\hsep{0.5cm}
\node[stagebox, fill=pipeblubg, draw=pipeblu, minimum width=4.8cm] (S1) at (0,0) {
  \begin{tabular}{@{}l@{}}
  \ifdefined\isarxiv\rule{0pt}{9pt}\\[-4pt]\fi
  {\small\bfseries\color{pipeblu} Dataset Construction}\\[4pt]
  {\footnotesize\color{pipeblu}\faCheck}~{\footnotesize Developer conversations}\\[1pt]
  {\footnotesize\color{pipeblu}\faCheck}~{\footnotesize Backed-out eval environment}\\[1pt]
  {\footnotesize\color{pipeblu}\faCheck}~{\footnotesize Candidate test retrieval}\\[1pt]
  {\footnotesize\color{pipeblu}\faCheck}~{\footnotesize Tests passing on master}
  \end{tabular}
};
\node[stagenum, fill=pipeblu] at ([xshift=8pt, yshift=-8pt]S1.north west) {1};
\node[stagebox, fill=pipeorgbg, draw=pipeorg, right=\hsep of S1, minimum width=3.6cm] (S2) {
  \begin{tabular}{@{}l@{}}
  \rule{0pt}{9pt}\\[-4pt]
  {\small\color{pipeorg}\faFilter}~{\small\bfseries\color{pipeorg} Prompt Quality}\\[4pt]
  {\footnotesize\color{pipeorg}\faCheck}~{\footnotesize Diff leakage removal}\\[1pt]
  {\footnotesize\color{pipeorg}\faCheck}~{\footnotesize Non-testable filtering}\\[1pt]
  {\footnotesize\color{pipeorg}\faCheck}~{\footnotesize Template exclusion}
  \end{tabular}
};
\node[stagenum, fill=pipeorg] at ([xshift=8pt, yshift=-8pt]S2.north west) {2};
\node[stagebox, fill=pipeorgbg, draw=pipeorg, right=\hsep of S2, minimum width=3.6cm] (S3) {
  \begin{tabular}{@{}l@{}}
  \rule{0pt}{9pt}\\[-4pt]
  {\small\color{pipeorg}\faFilter}~{\small\bfseries\color{pipeorg} Test Relevance}\\[4pt]
  {\footnotesize\color{pipeorg}\faCheck}~{\footnotesize Agentic classifier}\\[1pt]
  {\footnotesize\color{pipeorg}\faCheck}~{\footnotesize Code-path verification}\\[1pt]
  \phantom{{\footnotesize\faCheck}~{\footnotesize Placeholder}}
  \end{tabular}
};
\node[stagenum, fill=pipeorg] at ([xshift=8pt, yshift=-8pt]S3.north west) {3};
\node[stagebox, fill=pipegrnbg, draw=pipegrn, right=\hsep of S3, minimum width=3.6cm] (S4) {
  \begin{tabular}{@{}l@{}}
  \rule{0pt}{9pt}\\[-4pt]
  {\small\color{pipegrn}\faPlay}~{\small\bfseries\color{pipegrn} Test Validation}\\[4pt]
  {\footnotesize\color{pipegrn}\faCheck}~{\footnotesize Multi-run F2P check}\\[1pt]
  {\footnotesize\color{pipegrn}\faCheck}~{\footnotesize Flaky test exclusion}\\[1pt]
  \phantom{{\footnotesize\faCheck}~{\footnotesize Placeholder}}
  \end{tabular}
};
\node[stagenum, fill=pipegrn] at ([xshift=8pt, yshift=-8pt]S4.north west) {4};
\begin{scope}[on background layer]
  \node[rounded corners=10pt, dashed, draw=pipeorg!60, line width=0.8pt,
        inner xsep=8pt, inner ysep=12pt,
        fit=(S2)(S3)(S4),
        label={[font=\footnotesize\bfseries, text=pipeorg!80, anchor=south]above:{Verification~(\S\ref{subsec:verification})}}] {};
\end{scope}
\draw[sarrow] (S1.east) -- (S2.west);
\draw[sarrow] (S2.east) -- (S3.west);
\draw[sarrow] (S3.east) -- (S4.west);
\end{tikzpicture}
}
\caption{\normalfont \pipeline{} curation pipeline overview. Dataset construction assembles candidate tasks; \pipeline{}'s three verification filters, ordered so that semantic checks narrow the candidate set before execution, produce verified benchmark tasks.}
\label{fig:pipeline-overview}
\end{figure*}

\subsection{Task Formulation}
\label{subsec:task}

\paragraph{Agent Task} Each benchmark entry consists of three components: a verbatim developer prompt, a solution diff (the code change that landed in response), and a set of executable tests that provide a pass/fail evaluation signal.

\paragraph{Evaluation Environment} At evaluation time, the agent receives only the prompt; the solution diff is hidden, and the tests run post-hoc to verify the outcome. Tasks are interactive and the agent is expected to use the tools at its disposal to navigate the monorepo and produce a patch in response to the prompt. We do not employ any LLM judges~\cite{zhengJudgingLLMasajudgeMTbench2023}, relying exclusively on the executable tests to provide an evaluation signal.

\subsection{Dataset Construction}
\label{subsec:construction}

We build the benchmark in two stages. This section describes \emph{basic dataset construction}: assembling a candidate set of developer-agent trajectories, each paired with a backed-out solution diff and a permissive candidate test set. The candidate set is not yet a reliable evaluation signal, as we discuss below; Section~\ref{subsec:verification} describes the \emph{verification} stage that filters these candidates down to a verified subset admitted into the final benchmark.

\paragraph{Authentic Developer Conversations}

We source data from authentic conversations between developers and a production AI coding assistant. Each conversation is mapped to a solution diff that was subsequently approved, merged, and committed to the monorepo. The coding assistant is instrumented to log whenever AI-generated content is accepted by the developer and retained in the final committed diff. This \emph{AI provenance} metric enables us to track the connection between a conversation and its resulting landed diff, and to quantify the proportion of each diff that originates from AI suggestions. To maintain a strict one-to-one mapping between developer intent and code modifications, we exclude diffs that resulted from more than one developer-agent conversation. This helps ensure that the developer's request is not underspecified at the time of evaluation, enabling reliable evaluation of whether an agent's output matches the expected solution.

\paragraph{Backed-out Evaluation Environment}

As discussed in Section~\ref{sec:challenges}, monorepo infrastructure constraints can prevent reproducing the exact environment at a past commit. However, hiding the solution diff from the agent's environment helps ensure the agent does not ``cheat'' on a task by reading the committed solution files. We achieve this by \emph{backing out} the solution diff, that is, reverting the change on the current codebase while leaving all other code unchanged, and running the agent evaluation on top of this pre-change state. Due to the dynamic nature of a large codebase, some diffs cannot be cleanly backed out due to conflicting changes and are excluded from the benchmark. This eligibility constraint is time-sensitive: as the codebase evolves, an increasing fraction of historical diffs accumulates conflicts with the current master and falls out of eligibility, so freezing a benchmark snapshot would steadily erode its volume. \pipeline{} mitigates this by producing a \emph{rolling benchmark} that periodically re-curates its task set from developer-agent sessions whose solution diffs still back out cleanly against the current master. As a side benefit, this design also helps ensure the benchmark stays fresh, contamination-free, and in-distribution with current development practices.

\paragraph{Candidate Test Retrieval}

We employ probabilistic test retrieval, a build-system service that returns tests statistically likely to be affected by a given set of changed files, to surface a permissive candidate set of tests for each diff. The candidate set is restricted to tests that were last seen passing on master and that are not flagged as disabled or known-flaky. The natural next step would be to execute each candidate test on both the pre-change and post-change commits, admitting any test that fails before the change and passes after it as a fail-to-pass (F2P) label. This naive labeling procedure is unreliable in practice: transient infrastructure failures on the pre-change commit can cause unrelated tests to fail and then pass on the post-change commit, getting incorrectly admitted as F2P labels. The verification stage in Section~\ref{subsec:verification} addresses this by inserting prompt and test-relevance filters upstream of execution and applying a multi-run validation downstream.

\subsection{Verification}
\label{subsec:verification}

SWE-Bench Verified~\cite{IntroducingSWEbenchVerified} demonstrated that raw, automatically constructed task pools can misjudge agent performance: in that effort, 90 software engineers manually screened each task to remove three classes of noise: (i) unit tests so narrow or unrelated to the issue that they reject correct solutions, (ii) underspecified issue descriptions, and (iii) brittle development environments that cause tests to fail regardless of the solution. The raw candidate set from Section~\ref{subsec:construction} suffers from analogous failure modes: candidate tests with no causal connection to the diff, prompts that cannot be reliably evaluated by tests, and transient infrastructure failures that incorrectly admit flaky tests as F2P labels. The rolling re-curation cadence required by our setting (Section~\ref{subsec:construction}) makes per-cycle human verification infeasible at this volume, so we automate the verification step.

Verification produces a \emph{verified subset} of the candidate benchmark in three stages: prompt-quality filters that remove untestable or polluted prompts, a test relevance filter that reduces the candidate test set to those causally connected to the diff, and a multi-run execution check that admits the surviving tests as fail-to-pass (F2P) evaluation signal. These stages are deliberately ordered so that semantic filters narrow the candidate set before any tests are executed, so the multi-run F2P check runs only on a small, high-precision survivor set.

\vspace{1em}
\textbf{Prompt Quality Filters}

\paragraph{Solution Diff Leakage}
We identified cases where prompts referenced the solution diff. While the solution diff's code changes are hidden from the agent by backing out the change from the repository, agents retain access to internal knowledge tools (e.g., version control APIs, diff metadata) that cannot be restricted without breaking normal functionality. If the prompt explicitly references the solution diff by identifier, the agent can retrieve it through these tools, compromising evaluation integrity. For example, prompts like ``address the CI failure on diff D123'' can lead the agent to query the version control system and retrieve the solution. To prevent this, we remove any conversations that reference the solution diff in the developer's request.

\paragraph{Non-Testable Prompts.} We use an LLM-powered classifier to filter prompts based on whether they can be reliably evaluated with tests. The classifier is built on Claude Sonnet 4.5 and is invoked once per conversation on the first user message, truncated to 4000 characters; no diff or repository context is provided, since the goal is to judge testability from the developer's request alone. The classifier assigns one of seventeen task-type categories, partitioned a priori into a testable set (e.g., bug fix, feature request, refactoring) and an untestable set (e.g., documentation-only, code explanation, UI-only). A prompt is admitted only if the predicted category falls in the testable set.

\paragraph{Template Prompts}
We exclude conversations where the developer message matches known system prompts or template messages. These automated messages do not represent genuine devel\-oper-agent interactive coding tasks. Common patterns include conversation summarization requests, automated diff creation prompts, and merge conflict resolution templates.

\vspace{1em}
\textbf{Test-Quality Filters}
\paragraph{Test Relevance Filtering.} An agentic classifier verifies that for each (diff, test) pair, the test is affected by the code changes, either directly or indirectly through dependencies. The Test Relevance Agent is built on a frontier LLM (Claude Sonnet 4.5). For each diff, the agent receives the diff title, summary, and code changes alongside the full list of candidate tests. It then retrieves each candidate test's source from the monorepo via a code-search tool, judges relevance, and returns a filtered list of relevant tests. The agent enforces four relevance criteria, deliberately mirrored in the manual annotation rubric (Section~\ref{subsec:manual-verification}): a test is relevant if (i) the diff modifies the test or its fixtures, (ii) it directly exercises modified code, (iii) it exercises code that imports or calls modified code, or (iv) it resides in the same or a directly related test file. Naming similarity without a code dependency is explicitly insufficient. Because candidate test discovery is intentionally permissive, running execution-based F2P verification directly on every candidate is unreliable at this scale: transient infrastructure failures can cause unrelated tests to be incorrectly admitted as F2P signal. Filtering candidates by relevance first lets the downstream execution stage spend its multi-run budget on a small, high-precision survivor set rather than the full candidate pool.

\paragraph{Test Validation.} The relevance filter does not execute tests, so it cannot establish fail-to-pass behavior. Since candidate tests are restricted to those last seen passing on master (Section~\ref{subsec:construction}), the post-change side is taken as given. To produce the F2P signal, we execute each surviving candidate three times on the pre-change (backed-out) commit. The first two runs serve as warmup: in the distributed build environment, initial runs frequently fail due to sandbox provisioning rather than the code change itself. A test is admitted as fail-to-pass only if the third run fails. As an additional safeguard, the final test log is matched against known transient-failure patterns (e.g., sandbox errors, test-runner timeouts); matches are dropped rather than admitted as F2P.

\section{Evaluation}
\label{sec:evaluation}

We apply \pipeline{} to a single industrial monorepo, producing \benchmark{} from production developer-agent sessions. We first describe the resulting benchmark's composition and validate \pipeline{}'s automated classifiers and pipeline ordering. We then present two case studies demonstrating how the benchmark informs deployment decisions.

\paragraph{Scope.}
All evaluations in this section share two scope choices. First, we draw conversations from a single production AI coding assistant; the pipeline is agent-agnostic, but running against multiple agents is left to future work. Second, we restrict to single-turn conversations in which a developer provides one prompt and the agent produces code changes without intermediate human feedback. Both are case-study constraints rather than pipeline-level limitations; their implications are discussed in Section~\ref{sec:limitations}.

\subsection{Composition}
\label{subsec:distribution}

This section reports the composition of the \benchmark{} snapshot used in our case study, organized along three axes: language, task type, and per-task structural statistics. Together they support a quantitative comparison against existing public benchmarks (Table~\ref{tab:swe-bench-comparison}), evidencing how the curated benchmark differs from synthetic or open-source baselines.

\begin{figure}[t]
\centering
\includegraphics[width=0.9\columnwidth]{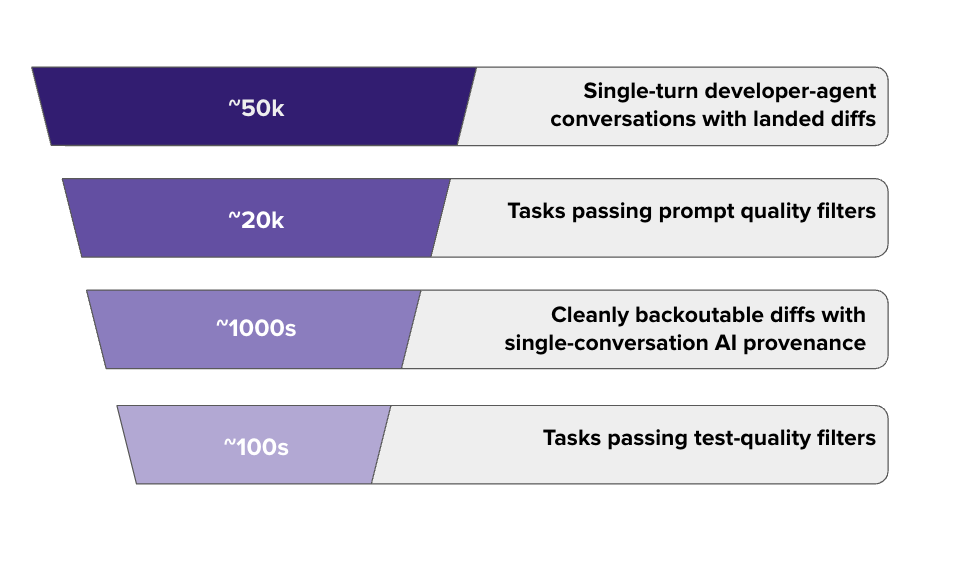}
\caption{\normalfont Data funnel for the case study, restricted to single-turn conversations: tens of thousands of candidate diffs are progressively filtered down to on the order of hundreds of benchmark tasks. Lightweight prompt-quality filters run before the more expensive backoutability and test discovery steps, reducing the input to costly stages and keeping end-to-end curation tractable at the rolling re-curation cadence.}
\label{fig:funnel}
\end{figure}

\begin{figure}[t]
\centering
\includegraphics[width=0.85\columnwidth]{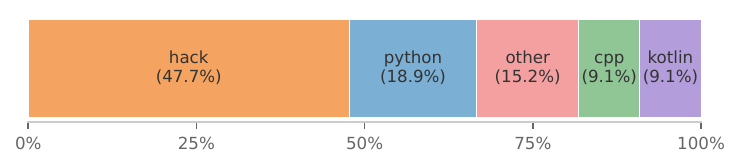}
\caption{\normalfont Language distribution of \benchmark{}. Hack accounts for the plurality of tasks, followed by Python, C++, Kotlin, and other languages, reflecting the polyglot nature of the host monorepo.}
\label{fig:language-distribution}
\end{figure}

\paragraph{Language Distribution.}
Figure~\ref{fig:language-distribution} shows the per-language share of tasks. The distribution is dominated by Hack, reflecting the language footprint of the host monorepo, and stands in contrast to public benchmarks: SWE-Bench and SWE-Bench Verified~\cite{jimenezSWEbenchCanLanguage2024,IntroducingSWEbenchVerified} are Python-only, and even multi-language extensions concentrate on widely available open-source ecosystems~\cite{rashidSWEPolyBenchMultilanguageBenchmark2025,zanMultiSWEbenchMultilingualBenchmark2025}.

\begin{figure}[t]
\centering
\includegraphics[width=0.95\columnwidth]{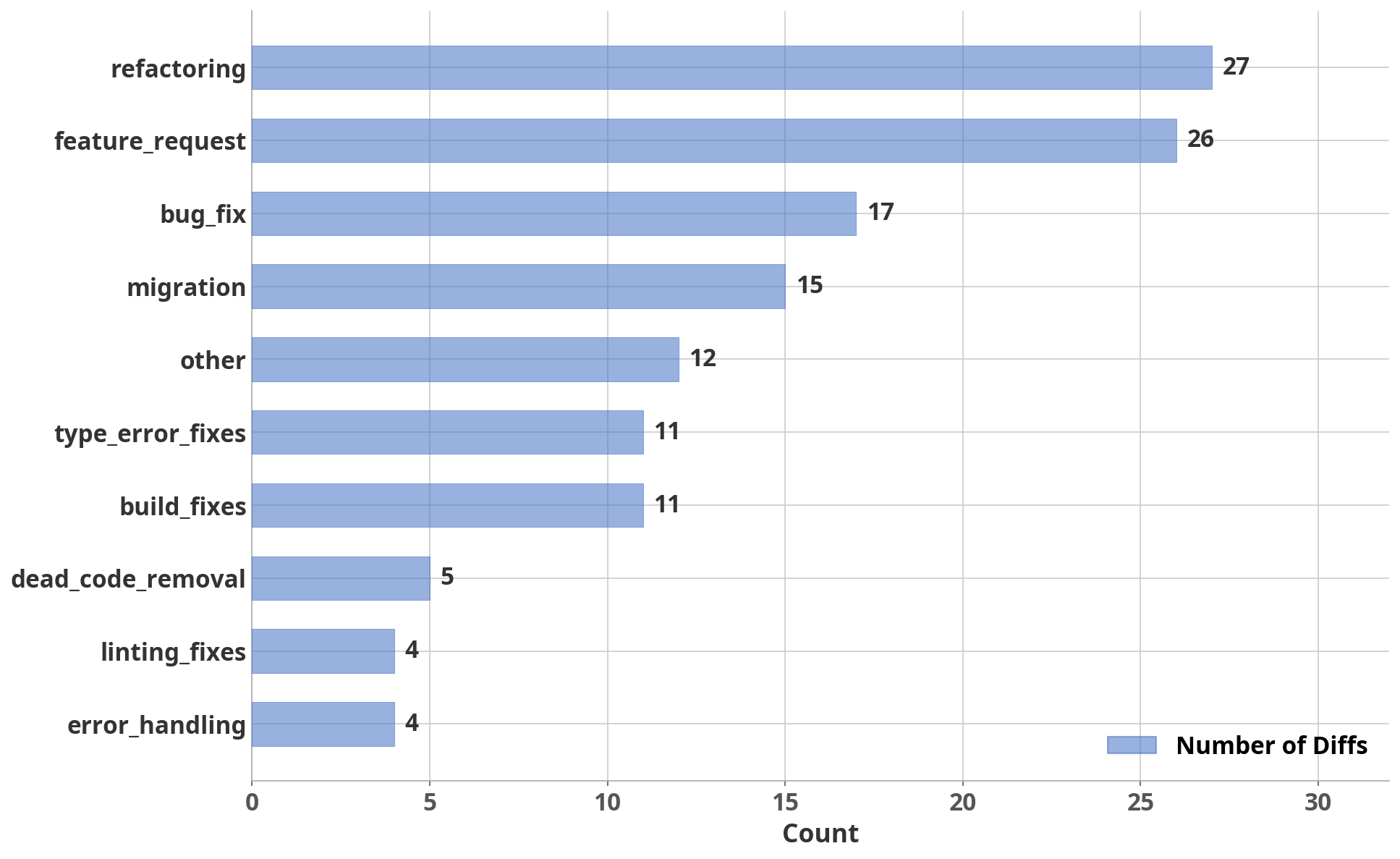}
\caption{\normalfont Tasks per category in \benchmark{}, assigned by an LLM-powered task classifier based on prompt content. No single category exceeds a quarter of the benchmark.}
\label{fig:task-categories}
\end{figure}

\paragraph{Task-type Distribution.}
Figure~\ref{fig:task-categories} shows the distribution of tasks by category, assigned by the LLM-powered task classifier (Section~\ref{subsec:verification}). No single category exceeds a quarter of the benchmark, with refactoring and feature requests leading and bug fixes representing a smaller share. This differs from issue-derived benchmarks where bug-fix tasks dominate.

\paragraph{Per-task Statistics.}
Table~\ref{tab:task-stats} summarizes per-task structural statistics. Median diff sizes are comparable to SWE-Bench~\cite{jimenezSWEbenchCanLanguage2024}, though the mean is skewed by a long right tail of large agent-authored diffs. The F2P signal is targeted: most tasks carry one or two fail-to-pass tests rather than a full test suite.

\begin{table}[t]
\centering
\caption{\normalfont Per-task structural statistics for \benchmark{}. Mean is reported alongside median to expose the right-tail skew driven by large agent-authored diffs.}
\label{tab:task-stats}
\begin{tabular}{lrrr}
\toprule
 & Mean & Median & Max \\
\midrule
Diff size (lines) & 697.8 & 39 & 77{,}879 \\
Files changed & 3.0 & 2 & 21 \\
Fail-to-pass tests & 3.3 & 1 & 39 \\
Total tests & 8.3 & 3 & 51 \\
\bottomrule
\end{tabular}
\end{table}

\paragraph{Prompt Realism.}
One contribution of \benchmark{} is capturing \textit{authentic} developer-agent interactions rather than synthetic task descriptions.
A qualitative review of all 132 verified prompts surfaced several emergent themes that set \benchmark{} apart from benchmarks sourced from GitHub issue bodies.
Compared to such benchmarks, these prompts more often
\begin{itemize}
\item \textbf{Assume implicit enterprise context.} Developers reference internal vocabulary (e.g., the internal name of a feature flag system), prior diffs by identifier, and domain-specific shorthand tied to internal tooling conventions, without ever defining them.
\item \textbf{Center on debugging from real breakage.} The prompt describes a regression introduced by a prior change and asks the agent to investigate root cause before proposing a fix (e.g., ``the power search filter is not working properly\ldots you can check this file and make it working the same way'').
\item \textbf{Require pattern following.} The prompt asks the agent to propagate changes by locating an existing example and applying it consistently throughout the codebase (e.g., ``find all the places that \texttt{anonymizedType} is used\ldots add a new type\ldots you can use the example of \texttt{one\_to\_one} or \texttt{group} to find all the call sites'').
\item \textbf{Embed IDE-workflow actions.} Prompts dispatch structured plan files, reference source files via absolute paths, or request version-control operations such as diff creation and reviewer assignment.
\end{itemize}
A majority of prompts exhibit at least one such pattern, and many exhibit multiple patterns simultaneously.
Most of these interaction modes are structurally absent from benchmarks derived from GitHub issue bodies, which tend to be self-contained and symptom-focused rather than context-dependent and action-oriented.

\begin{table*}[t]
\centering
\small
\caption{\normalfont Comparison of \benchmark{} against SWE-Bench~\cite{jimenezSWEbenchCanLanguage2024}. Both benchmarks evaluate agents at the repository level using fail-to-pass tests (no LLM judges), but diverge on provenance, scope, and sustainability.}
\label{tab:swe-bench-comparison}
\setlength{\tabcolsep}{5pt}
\renewcommand{\arraystretch}{1.25}
\begin{tabular}{@{}p{0.14\textwidth} p{0.38\textwidth} p{0.38\textwidth}@{}}
\toprule
\textbf{Axis} & \textbf{\benchmark{}} & \textbf{SWE-Bench} \\
\midrule
Prompt source
  & Verbatim prompts typed \emph{for an AI agent}
  & GitHub issue descriptions written \emph{for human developers} \\
Codebase scope
  & Industrial monorepo (ephemeral tooling, distributed builds)
  & 12 open-source GitHub repositories \\
Language coverage
  & 5+ languages, incl.\ low-resource and industrially relevant
  & Python only \\
Curation method
  & LLM classifiers for testability and test relevance (automated)
  & 90 engineers screen for test quality (Verified subset) \\
Contamination risk
  & Rolling re-curation; private codebase unseen by models
  & Frozen snapshot; public repos visible to model training \\
Task count
  & ${\sim}$130 (comparable to HumanEval~\cite{chenEvaluatingLargeLanguage2021}; automated)
  & 500 Verified~\cite{IntroducingSWEbenchVerified} (from 2{,}294 after 90-engineer screening) \\
\bottomrule
\end{tabular}
\end{table*}
\subsection{Empirical Validation of Automatic Curation}
\label{subsec:manual-verification}

To validate the LLM-powered classifiers, two annotators (the first two authors, each with 5+ years of SE experience) independently judged samples following the same protocol as the classifier: assign a fine-grained task type from the seventeen-category taxonomy (Section~\ref{subsec:verification}), then derive binary testability from the type's partition membership. Disagreements were resolved by discussion.

\subsubsection{Task Classification Accuracy}
\label{sec:task-classification-accuracy}

We sampled \taskClassifTotal{} tasks---50 classified as testable and 50 as non-testable by the classifier. Human consensus relabeled them as \taskClassifConsensusTestable{} testable and \taskClassifConsensusNontestable{} non-testable.
The classifier achieved \taskClassifAcc{} agreement with consensus (\taskClassifAgree/\taskClassifTotal{}, $\kappa = \taskClassifKappa$, substantial agreement), producing \taskClassifFP{} false positives and \taskClassifFN{} false negatives (precision \taskClassifPrec{}, recall \taskClassifRec{}, F$_1$ \taskClassifFone{}).
For reference, the two annotators agreed in \taskClassifIAA{} of cases ($\kappa = \taskClassifIAAKappa$) \emph{before} discussion, confirming inherent subjectivity and that the classifier operates near human-level reliability.

The \taskClassifFN{} false negatives fell into two patterns:
(i)~implicit code-change requests classified as \emph{debug investigation} (5~cases), where prompts like ``Can you add some logs'' were treated as investigation rather than a code change;
and (ii)~category misalignment (5~cases), where the classifier assigned non-testable categories (e.g., \emph{documentation only}, \emph{UI only}) to tasks annotators deemed testable---notably, a SKILL file change (a Markdown file consumed by coding agents) was labeled \emph{documentation only} despite affecting agent execution.
The \taskClassifFP{} false positives involved labeling ambiguous prompts as testable via \emph{other} or \emph{bug fix} when the task had no clear expected outcome.

\begin{insight}
\pipeline{}'s task classifier is effective: at \taskClassifAcc{} accuracy it matches human inter-annotator reliability (\taskClassifIAA{}), and when it errs it tends to discard valid tasks rather than admit untestable ones---shrinking the benchmark slightly but keeping evaluation scores reliable.
\end{insight}

\subsubsection{Test Relevance Accuracy}
\label{sec:test-relevance-accuracy}

We sampled \testRelevanceTotal{} (diff, test) pairs---25 classified as relevant and 25 as non-relevant by the classifier---spanning \testRelevanceDiffs{} unique diffs. Annotators judged relevance against four criteria: (1)~the diff modifies the test or its fixtures; (2)~the test directly exercises the modified function; (3)~the test exercises code depending on the modified module; (4)~the test resides in the same module and exercises its functionality.

The classifier achieved \testRelevanceAcc{} agreement with consensus (\testRelevanceAgree/\testRelevanceTotal{}, $\kappa = \testRelevanceKappa$, substantial agreement), producing \testRelevanceFN{} false negatives and \testRelevanceFP{} false positives (precision~\testRelevancePrec{}, recall~\testRelevanceRec{}, F$_1$~\testRelevanceFone{}).
The two annotators agreed in only \testRelevanceIAA{} of cases ($\kappa = \testRelevanceIAAKappa$, moderate), reflecting greater subjectivity---particularly around criterion~4. The classifier's \testRelevanceAcc{} accuracy exceeds raw inter-annotator agreement, suggesting it captures a more consistent signal than either individual annotator.

The \testRelevanceFN{} false negatives involved same-module tests that did not exercise the specific modified function (2~cases), an indirect build dependency the classifier could not trace (1~case), and an allowlist pattern test (1~case).
The \testRelevanceFP{} false positives were caused by name similarity without actual dependency chains (2~cases) and a subtle class-name difference that also confused one annotator (1~case).

\begin{insight}
\pipeline{}'s test-relevance classifier achieves \testRelevanceAcc{} accuracy, exceeding the \testRelevanceIAA{} raw inter-annotator agreement, with near-symmetric errors (\testRelevanceFN{} FN vs.\ \testRelevanceFP{} FP), reliably identifying tests that exercise the changed code.
\end{insight}

\subsection{Pipeline Stage Ordering}
\label{subsec:ordering-audit}

\paragraph{Motivation.} We initially built \benchmark{} execution-first: discover candidate tests, run a single-pass F2P check at the base commit, then apply verification. When running at scale in a distributed build environment, single-pass execution is unreliable: tests can fail for reasons unrelated to the code change, including infrastructure errors, timeouts, and out-of-memory crashes. These spurious failures appear as F2P but are not caused by the code change; at evaluation time the same tests pass even with an empty patch, inflating solve rates. Multi-run execution filters out such noise, but running it on the full unfiltered candidate set is prohibitively slow and expensive. Relevance-first reorders the pipeline so that the test-relevance filter narrows the candidate set semantically before any execution, and multi-run F2P is then applied only to the surviving set. For the same reason, the prompt-quality classifier is also moved upstream of backout and test discovery, so that only testable prompts enter the expensive stages (Figure~\ref{fig:funnel}).

\paragraph{Empirical Comparison.} We compare two pipeline orderings on the same source pool: (A)~\textit{execution-first}, which uses single-pass F2P detection to narrow the candidate set and then applies test-relevance filtering to the F2P subset; and (B)~\textit{relevance-first}, which uses test-relevance filtering directly to narrow the candidate set, skipping single-pass execution entirely. Both orderings finish with multi-run F2P detection, but they differ in what does the heavy lifting upstream. Relevance-first (B) retains \AuditRetentionPct{}\% of execution-first (A)'s F2P pairs (\AuditOverlapN{} of \AuditAN{}), admits \AuditNetGainPct{}\% more tasks net (\AuditBN{} vs \AuditAN{}), and broadens language coverage to include Kotlin, Swift, and TypeScript tasks absent from A. Per-category retention ranges from 100\% on type, build, and linting fixes to 60\% on dead-code removal. Per-diff signal strength is preserved: median F2P tests per diff is 1 under both variants.
\begin{insight}
Relevance-first ordering yields a +\AuditNetGainPct{}\% net gain over execution-first (\AuditBN{} vs \AuditAN{} tasks) and broadens language coverage, leading to a more balanced benchmark.
\end{insight}

\begin{figure}[t]
\centering
\includegraphics[width=0.9\columnwidth]{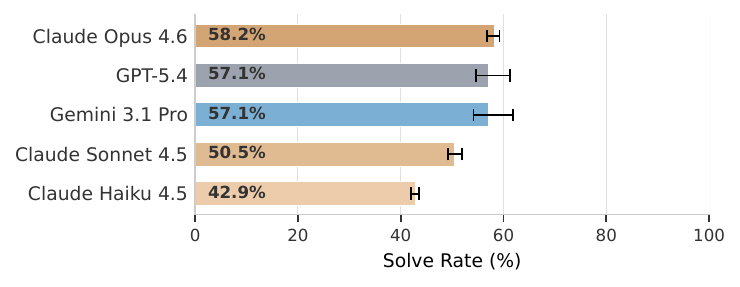}
\caption{\normalfont Model solve rates on \benchmark{} across five frontier models evaluated on 132 tasks with three independent runs each. Bar height shows the mean solve rate across runs; whiskers indicate the min and max of the three per-run solve rates, capturing run-to-run reproducibility.}
\label{fig:solve-rates}
\end{figure}

\begin{figure}[t]
\centering
\includegraphics[width=0.9\columnwidth]{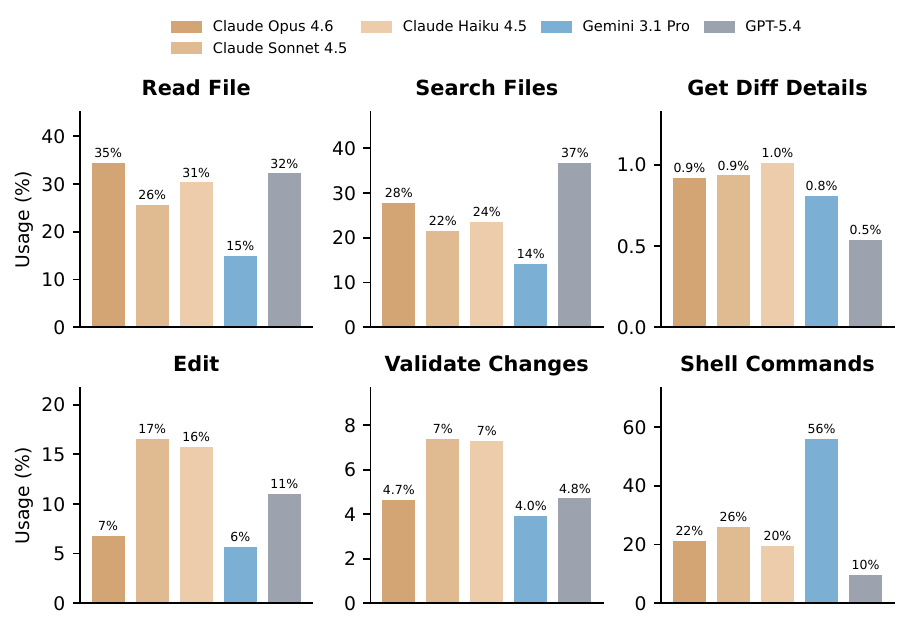}
\caption{\normalfont Tool usage distribution across five frontier models on \benchmark{}, shown as percentage of total tool calls per model.}
\label{fig:tool-usage}
\end{figure}

\subsection{Case Study 1: Model Comparison}
Next, we present a case study that demonstrates how a production-derived benchmark informs deployment decisions. We run the model evaluations on the F2P subset of the benchmark, i.e., only on diffs that have at least one fail-to-pass test.

\begin{figure*}[t]
\centering
\includegraphics[width=\textwidth]{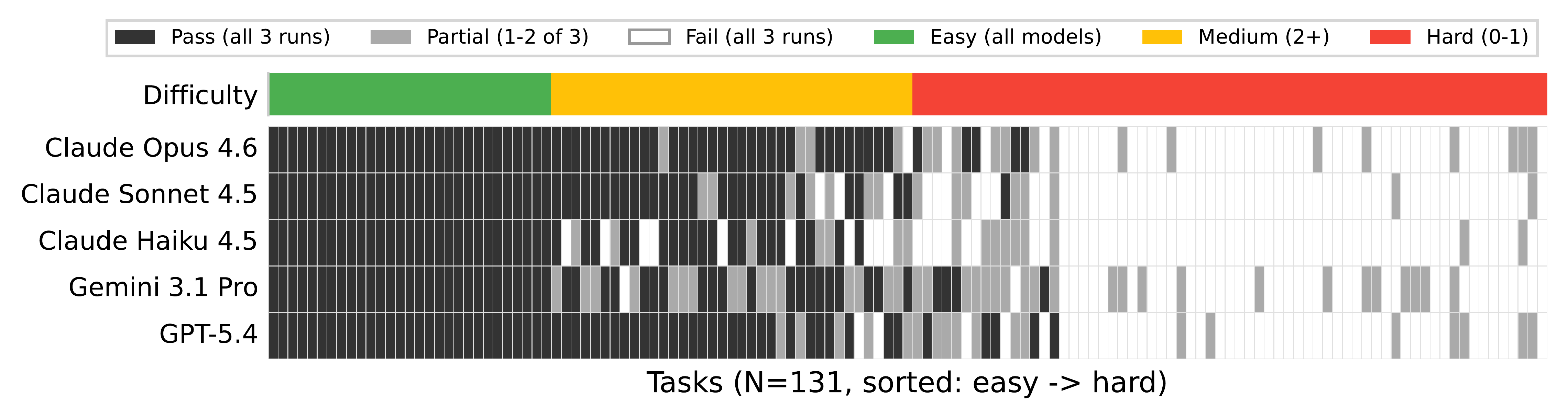}
\caption{\normalfont Per-diff consistency heatmap across five frontier models on \benchmark{}. Models are grouped by provider (Claude Opus 4.6, Claude Sonnet 4.5, Claude Haiku 4.5, Gemini 3.1 Pro, GPT-5.4, top to bottom); see Figure~\ref{fig:solve-rates} for solve-rate ranking. Diffs are sorted by difficulty (easiest on left, hardest on right). Cell colors indicate pass consistency across three independent runs: black = pass all 3 runs (consistently solvable), gray = pass 1--2 of 3 runs (non-deterministic), white = fail all 3 runs (consistently unsolvable). The top strip shows task difficulty: green = all models pass all runs, yellow = 2--4 models pass strictly, red = 0--1 models pass strictly.}
\label{fig:heatmap}
\end{figure*}

We evaluated five frontier models (Claude Opus 4.6, Claude Sonnet 4.5, Claude Haiku 4.5, Gemini 3.1 Pro, and GPT-5.4) on a \benchmark{} snapshot of 132 F2P tasks. Each model was evaluated three times independently to measure run-to-run variance. Figure~\ref{fig:solve-rates} presents the solve rates with min/max whiskers across the three runs.
The evaluation results reveal performance tiers rather than a fine-grained ranking: Claude Opus 4.6, GPT-5.4, and Gemini 3.1 Pro cluster within roughly one percentage point of each other with overlapping min/max ranges, while Claude Sonnet 4.5 and Claude Haiku 4.5 trail by clear margins. Given three runs per model on 132 tasks, differences within the top cluster fall within run-to-run variance; the reliable deployment signal is the separation between tiers, not the ordering inside the top cluster.
Figure~\ref{fig:tool-usage} shows the tool usage distribution across models. Models at similar solve rates adopt notably different strategies: for example, Gemini 3.1 Pro relies heavily on shell commands (56\% of tool calls) while GPT-5.4 favors search (37\%).
Figure~\ref{fig:heatmap} provides a per-diff consistency view: diffs are sorted by difficulty (easiest on the left, hardest on the right), and each cell indicates whether a model passed all three runs (black), one or two runs (gray), or none (white). A difficulty sidebar at the top categorizes each diff as easy (all models pass strictly), medium (2--4 models pass), or hard (0--1 models pass). The heatmap reveals that a substantial fraction of diffs are consistently solved or consistently failed, while a smaller set exhibits non-deterministic behavior across runs.

\begin{insight}
\benchmark{} separates frontier models into clear performance tiers, providing a quantitative signal for deployment decisions; differences within the top tier fall inside run-to-run variance and should not, on their own, drive model selection. The heatmap confirms stable easy/hard task clusters rather than random noise. Tool usage patterns reveal distinct strategies: models at similar solve rates rely on different tool mixes to reach comparable outcomes.
\end{insight}

\subsection{Case Study 2: Harness Engineering}
We compare two evaluation harnesses on \benchmark{}. \textbf{Basic Agent} uses a limited toolset and simpler search tools. \textbf{Advanced Agent} features a comprehensive toolset with prompts designed to generalize across model families. The Advanced Agent harness has roughly 3x more tools than Basic Agent, including code navigation, local diagnostics, formatting, and knowledge search. Additionally, we evaluate the effect of \textbf{Context Files}, which are developer-authored Markdown files that encode internal codebase knowledge such as coding conventions, build system details, and project-specific patterns. These files can be loaded at evaluation time to help the agent navigate the proprietary codebase.

\begin{figure}[t]
\centering
\includegraphics[width=0.81\columnwidth]{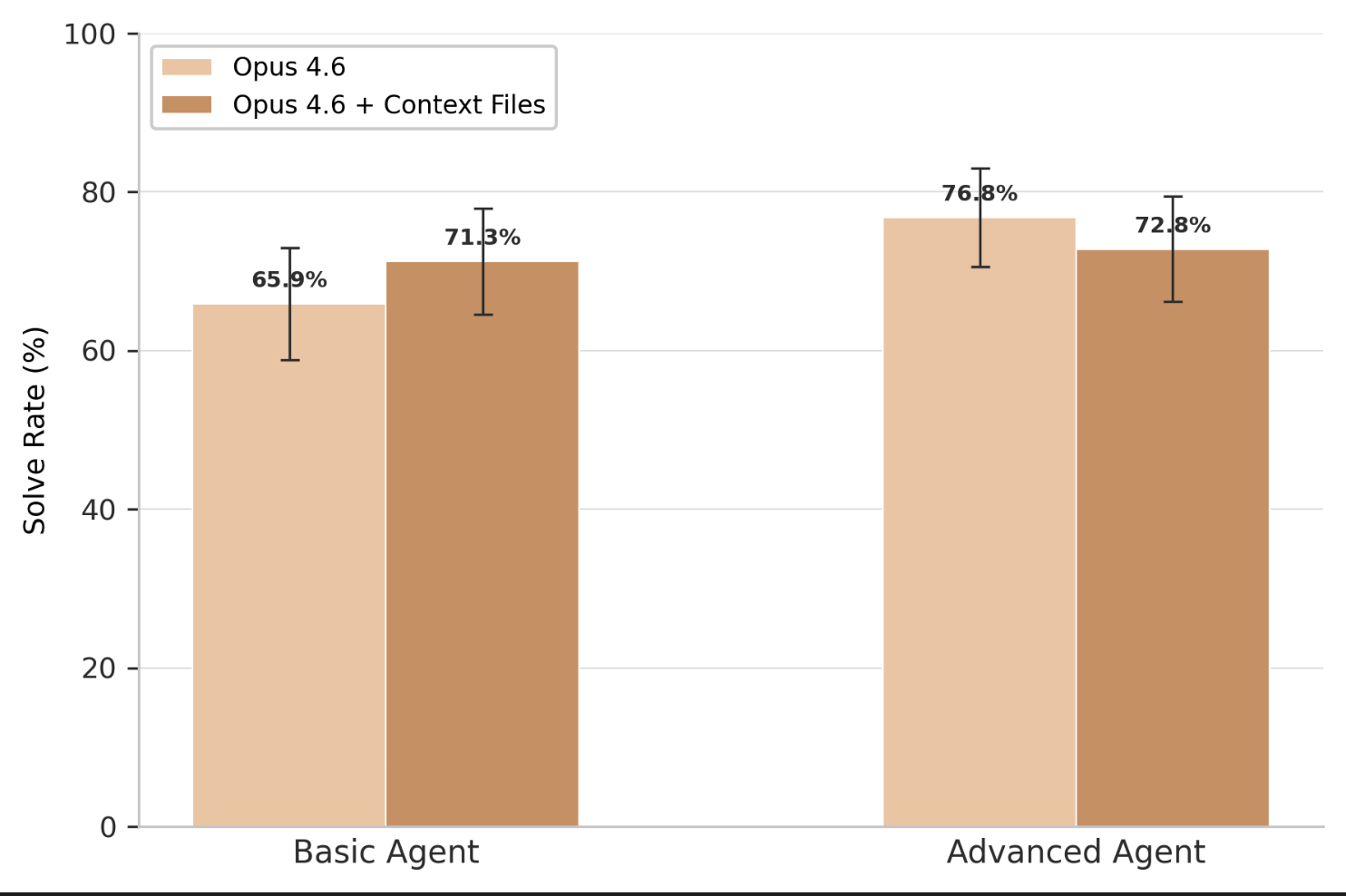}
\caption{\normalfont Harness evaluation comparison on \benchmark{}. Advanced Agent consistently outperforms Basic Agent. Basic Agent's limited search tools lead to prolonged search operations (see Table~\ref{tab:tool_usage}). Advanced Agent provides better code navigation, leading to fewer searches and higher solve rates.}
\label{fig:harness-eval}
\end{figure}

Figure~\ref{fig:harness-eval} presents the results. The Advanced Agent harness consistently outperforms Basic Agent across configurations. The addition of Context Files lifts the solve rate on the Basic Agent harness (+6.4\,pp), where the improvement is driven by three mechanisms: Context Files surface domain-specific tool calls (e.g., targeted test runners and schema lookups) that do not appear in runs without them, they reduce the number of sessions that exhaust the maximum iteration budget (4 vs.\ 7), and they help the model converge on a solution despite weaker search tools. The cost is latency: Context File runs are roughly 32\% slower due to additional tool calls and context processing.

On the Advanced Agent harness, however, Context Files provide no benefit and can slightly hurt performance. The stronger harness already supplies the navigation and diagnostic capabilities that Context Files encode, so loading them adds context noise without new information. Without Context Files, the model spends more of its budget on its core loop (edit, validate, iterate) rather than processing redundant instructions.

\begin{insight}
Richer tooling lifts solve rates, and Context Files amplify weak harnesses by filling capability gaps. On strong harnesses the same files can become redundant context that slightly hurts performance.
\end{insight}

\begin{table}[t]
\centering
\caption{\normalfont Tool usage distribution (\%) across harness configurations (Opus 4.6). Basic Agent allocates significantly more calls to search, while Advanced Agent distributes effort more evenly across navigation, planning, and validation.}
\label{tab:tool_usage}
\small
\begin{tabular}{lcc}
\toprule
Tool & Basic Agent & Advanced Agent \\
\midrule
Read File & 36.8 & 26.5 \\
Search Files & 31.4 & 13.4 \\
Shell & 17.0 & 6.1 \\
Todo Write & --- & 13.7 \\
Edit & 9.1 & --- \\
Str Replace Edit & --- & 9.4 \\
Validate Changes & 4.0 & 6.3 \\
\bottomrule
\end{tabular}
\end{table}

\section{Related Work}
\label{sec:related}

\subsection{Benchmarks for Code Generation and Software Engineering Agents}

Evaluation of AI-assisted code generation has progressed through two distinct phases. The first generation of benchmarks, including HumanEval~\cite{chenEvaluatingLargeLanguage2021}, MBPP~\cite{austinProgramSynthesisLarge2021}, APPS~\cite{hendrycksMeasuringCodingChallenge2021}, and CodeContests~\cite{liCompetitionlevelCodeGeneration2022}, evaluates the ability of large language models to synthesize standalone functions from natural language specifications. These benchmarks measure \emph{functional correctness}, whether generated code passes a suite of unit tests, but abstract away the complexity inherent in professional software development: navigating large codebases, resolving cross-file dependencies, and integrating changes into existing systems.

The second generation addresses these limitations by situating evaluation within realistic repository contexts. SWE-bench~\cite{jimenezSWEbenchCanLanguage2024} pioneered this approach by constructing tasks from real GitHub issues paired with their corresponding pull request solutions, using the repository's test suite as an oracle. This benchmark, along with its human-validated subset SWE-bench Verified~\cite{IntroducingSWEbenchVerified} and training-oriented variant SWE-Gym~\cite{panTrainingSoftwareEngineering2025}, has become the primary evaluation standard for software engineering agents. Subsequent work has extended this paradigm to additional programming languages: SWE-Sharp-Bench~\cite{mhatreSWESharpBenchReproducibleBenchmark2025} targets C\#, SWE-PolyBench~\cite{rashidSWEPolyBenchMultilanguageBenchmark2025} covers Java, JavaScript, and TypeScript, and Multi-SWE-bench~\cite{zanMultiSWEbenchMultilingualBenchmark2025} spans seven languages. CrossCodeEval~\cite{dingCrossCodeEvalDiverseMultilingual2023} further evaluates cross-file code completion capabilities.

\benchmark{} shares the repository-level evaluation paradigm with SWE-bench but differs in a key aspect: the source of natural language task specifications. Existing benchmarks derive their prompts from GitHub issue descriptions, text written to communicate with other developers, not to instruct an AI system. In contrast, \benchmark{} captures the \emph{verbatim prompts} that developers typed into an AI coding assistant. This distinction matters because issue descriptions are written for a human audience and tend to describe symptoms rather than desired behavior, whereas prompts to AI assistants are directive and action-oriented, often embedding implicit enterprise context and IDE-specific workflow actions (Section~\ref{subsec:distribution}).

\subsection{Developer Interactions with AI Coding Assistants}

Understanding how developers actually interact with AI coding tools helps construct representative benchmarks. Several empirical studies have investigated these interaction patterns. Barke et al.~\cite{barkeGroundedCopilotHow2023} conducted an observational study of 20 programmers using GitHub Copilot and identified two primary interaction modes: \emph{acceleration mode}, where developers have a clear goal and use the AI to implement it faster, and \emph{exploration mode}, where developers are uncertain and use the AI to discover possible approaches. Vaithilingam et al.~\cite{vaithilingamExpectationVsExperience2022} found that while Copilot did not consistently improve task completion time, developers valued it for providing useful starting points and reducing the need to search documentation. Studies of developer trust~\cite{brownIdentifyingFactorsThat2024} have shown that acceptance of AI suggestions depends on factors including suggestion quality, the developer's expertise in the relevant language, and the development context (e.g., production code versus tests).

Other work has examined how developers communicate with conversational AI assistants. Xiao et al.~\cite{xiaoDevGPTStudyingDeveloperChatGPT2024} curated a dataset of developer conversations with ChatGPT shared on GitHub, enabling analysis of real-world usage patterns. Khojah et al.~\cite{khojahHumantoHumanHumantoBotConversations2024} compared conversations between developers and those between developers and AI assistants, finding fundamental differences in structure and content. Mozannar et al.~\cite{mozannarReadingLinesModeling2024} developed a model of user behavior when interacting with code completion systems, quantifying the time costs of different interaction patterns.

These studies reveal a gap that \benchmark{} addresses: while we understand increasingly more about how developers interact with AI tools, existing benchmarks do not capture these authentic interactions. By sourcing tasks from actual usage of an AI coding assistant, \benchmark{} provides evaluation data that reflects real developer intent and communication patterns, complementing insights from observational studies with a benchmark grounded in authentic use.

\subsection{Benchmark Quality and Reinforcement Learning for Code}

The reliability of benchmark-driven evaluation depends on data quality. Two challenges are particularly relevant: test flakiness and data contamination. Flaky tests, tests that produce non-deterministic outcomes without changes to the code under test, can introduce noise into evaluation metrics~\cite{parrySurveyFlakyTests2021}. \benchmark{} mitigates this through multi-run execution with warmup runs that absorb transient infrastructure failures before a final decisive run establishes the F2P signal. Data contamination occurs when benchmark data appears in model training corpora, leading to inflated performance estimates. LiveCodeBench~\cite{jainLiveCodeBenchHolisticContamination2024} addresses this by continuously collecting new problems from programming competitions, ensuring temporal separation from training data. \benchmark{} adopts a similar \emph{rolling benchmark} design, periodically refreshing the task set with recent samples that post-date model training cutoffs.

Beyond evaluation, execution-based signals from test suites can guide model improvement. Research in automated program repair (APR) has demonstrated that large language models, when combined with test feedback, can substantially outperform traditional template-based repair techniques~\cite{xiaAutomatedProgramRepair2023}. Conversational approaches such as ChatRepair~\cite{xiaAutomatedProgramRepair2024} further improve repair effectiveness by iteratively refining patches based on test failure information. This principle extends to model training: reinforcement learning methods can use test outcomes as reward signals. CodeRL~\cite{leCodeRLMasteringCode2022} introduced an actor-critic framework where a critic network predicts functional correctness to provide training signal, while RLEF~\cite{gehringRLEFGroundingCode2025} demonstrated that models trained to leverage execution feedback require an order of magnitude fewer samples to achieve comparable performance. The current case study yields hundreds of tasks rather than the thousands typically required for RL training, but the execution-based fail-to-pass signal produced by \pipeline{} is structurally compatible with reward signal construction as the data funnel scales.

\section{Threats to Validity}
\label{sec:limitations}

\paragraph{Construct Validity.}
The multi-stage filtering pipeline reduces the initial corpus to a few hundred benchmark instances in the current case study, trading data volume for quality. Widening the data funnel (e.g., multiple agents, multi-turn conversations) would increase volume; at present, the yield is below the scale typically required for reward-signal-based post-training via reinforcement learning (RL). The current case study is scoped to single-turn interactions; extending \pipeline{} to multi-turn conversations would require revisiting the task classifier to handle iterative prompts, though downstream pipeline stages (dataset construction, test relevance filtering, test validation) are agnostic to conversation structure. A separate concern is that \pipeline{} defines correctness as passing the curated F2P tests. The relevance filter (Section~\ref{subsec:verification}) and multi-run validation help ensure these tests genuinely exercise the diff, but test-pass remains a necessary, not sufficient, correctness signal: a model could in principle satisfy the tests through narrow patches or test-specific workarounds without addressing the developer's broader intent. We expect this risk to be small in practice given that the relevance filter grounds each test in the changed code, but ruling it out fully would require complementary judgments such as code review or behavioral comparison against the human-authored solution.

\paragraph{Internal Validity.}
The evaluation harness exposes a set of tools to the agent under evaluation, some of which may inadvertently disclose privileged information. In particular, a tool that retrieves the current local diff could expose the \emph{backout diff}, the inverse patch that would revert the committed change, effectively leaking the ground-truth solution and compromising evaluation integrity.

\paragraph{External Validity.}
State-of-the-art models currently achieve a solve rate of approximately \SolveRateHighest{}\%, indicating that a fraction of benchmark instances may be insufficiently challenging. This ceiling effect limits the benchmark's discriminative power for distinguishing among frontier models. The aggressive verification pipeline retains on the order of hundreds of tasks from a starting pool of tens of thousands of candidate diffs, so \benchmark{} should be interpreted as a high-confidence \emph{sample} of testable production work rather than a complete characterization of all developer-agent interactions. Findings about model performance, harness effects, and tool usage are anchored to this filtered subset; task categories that the filters systematically exclude (e.g., UI-only changes, documentation, test generation, and multi-turn debugging) are out of scope and require complementary evaluation. The funnel also drops ambiguous or underspecified requests, interactions that did not culminate in a landed diff, and tasks without stable fail-to-pass labels; reported solve rates therefore measure performance on the verified, testable slice of production work---likely skewed toward cleaner, better-specified tasks---rather than on production traffic at large.

The source conversations come from a single production AI coding assistant. Developer phrasing may therefore reflect that assistant's UX and conversational affordances, and prompt distributions could shift if the same pipeline were run against a different agent. The pipeline is agent-agnostic by construction, so this is a coverage limitation of the present case study rather than a methodology constraint.

The LLM-based relevance filter (Section~\ref{subsec:ordering-audit}) has known blind spots for certain task categories. Dead-code-removal tasks are under-retained ($\sim$60\%), reflecting that deletion-style diffs rarely have a corresponding fail-to-pass test by construction, and feature-request tasks are mildly under-retained ($\sim$84\%), likely because new-feature tests in newly added test files are harder for the filter to ground against the diff. Findings about model performance on these task categories should be interpreted with this selection effect in mind.

\paragraph{Reliability.}
To mitigate threats to reliability, \pipeline{} employs multi-run execution with warmup runs that absorb transient infrastructure failures, admitting only tests whose final run produces a stable F2P signal. Each model is evaluated three times to quantify variance in reported solve rates. The rolling benchmark design, in which stale samples are periodically replaced with fresh ones, does not affect the primary use case of comparing models or harnesses: all candidates are evaluated on the same snapshot. The rolling design also confers advantages that frozen benchmarks lack: each refresh yields tasks that post-date model training cutoffs, providing a natural defense against data contamination, and ensures every task remains executable against the current codebase. Longitudinal comparisons across snapshots require controlling for task-set differences, but this is straightforward when a stable subset of tasks persists across refreshes.

\subsection{Planned Improvements}

Future work will address these threats along several dimensions. To improve pipeline throughput, we plan to parallelize base-commit test execution and introduce caching of backout-diff results, reducing end-to-end benchmark construction time. To improve environment faithfulness, we will incorporate recorded workspace snapshots to restore the exact set of local, uncommitted changes visible to the developer at authoring time.
\section{Conclusion}
\label{sec:conclusion}

This paper presented \pipeline{}, a replicable methodology for curating production-derived benchmarks for AI coding agents.
Our automated curation pipeline, including LLM-based task classification, test relevance verification, and multi-run stability checks, addresses the practical challenges of constructing reliable evaluation signals from monorepo environments.

Through \benchmark{}, a case study applying the methodology to one industrial monorepo, we demonstrated its utility: evaluation of five frontier models revealed solve rates ranging from \SolveRateLowest{}\% to \SolveRateHighest{}\%. Harness comparison experiments showed that stronger toolsets lift solve rates, and that developer-authored context files amplify weaker harnesses by filling capability gaps.

The methodology is designed to be replicable: organizations can follow this approach to construct analogous benchmarks from their own production data, tailored to their languages, conventions, and deployment requirements.

\ifdefined\isarxiv\else
\section*{Data Availability Statement}
\label{sec:das}

The \benchmark{} dataset cannot be released because it contains proprietary source code and internal developer prompts. The evaluation harness depends on internal build infrastructure that cannot be reproduced externally. To support adoption of the methodology, we release open-source, runnable implementations of the curation classifiers and manual annotation protocols at \url{https://anonymous.4open.science/r/REAP_artifact}. The release includes: (1) the task-type classifier that categorizes developer prompts and determines testability, (2) the test-relevance agent that clones a repository, explores it via shell commands, and classifies whether each candidate test is relevant to a given code change, and (3) the human-annotation guidelines used to validate both classifiers. These artifacts are sufficient for practitioners to re-implement the \pipeline{} curation pipeline against their own codebase and to audit the decisions driving each filtering stage.
\fi